# An analytical method to quantify the statistics of energy landscapes in random solid solutions

Ritesh Jagatramka    Chu Wang    Matthew Daly[*]

Department of Civil, Materials, and Environmental Engineering, University of Illinois at Chicago – 842 W. Taylor St., 2095 ERF (MC 246), Chicago, IL, 60607, United States

**ABSTRACT**

Recent studies of concentrated solid solutions have highlighted the role of varied solute interactions in the determination of a wide variety of mesoscale properties. These solute interactions emerge as spatial fluctuations in potential energy, which arise from local variations in the chemical environment. Although observations of potential energy fluctuations are well documented in the literature, there remains a paucity of methods to determine their statistics. Here, we present a set of analytical equations to quantify the statistics of potential energy landscapes in randomly arranged solid solutions. Our approach is based on a reparameterization of the relations of the embedded atom method in terms of the solute coordination environment. The final equations are general and can be applied to different crystal lattices and energy landscapes, provided the systems of interest can be described by sets of coordination relations. We leverage these statistical relations to study the cohesive energy and generalized planar fault energy landscapes of several different solid solutions. Analytical predictions are validated using molecular statics simulations, which find excellent agreement in most cases. The outcomes of this analysis provide new insights into phase stability and the interpretation of 'local' planar fault energies in solid solutions, which are topics of ongoing discussion within

---

[*]Corresponding author: mattdaly@uic.edu (M. Daly)

---

**1. INTRODUCTION**

Recent interest in concentrated systems has spurred research activity in the metallurgy of multicomponent solid solutions. These efforts have produced new physical insights into the role of solute interactions, with a focus on behaviors that emerge in concentrated mixtures. Studies in high entropy alloys (HEA)s, which are considered here as a subset of concentrated solid solutions, are perhaps the most visible. Notable examples include investigations of solute-induced lattice distortion [1–5], variations in dislocation glide barriers [6–13], and solute effects on planar and twinning fault energies [14–18]. While the effects of solute can be revealed by computational studies and inferred by experiments (e.g. as in Refs. [19–21]), new theoretical challenges have arisen concerning the prediction of material properties. For instance, traditional solute theories such as those for solid solution strengthening [22–25], find limited applicability in concentrated systems [8,13]. In many cases, the shortcomings of existing theories can be traced to the irregular nanoscale fluctuations in the potential energy (PE) landscape. These topology-induced PE fluctuations emerge from local variations in composition and solute pairings between constituents, whose effect becomes significant in concentrated solid solutions. Perhaps the most basic example of such a PE landscape is that of cohesive energy, but more complicated landscapes such as that of the Peierls barrier and the generalized planar fault energy (GPFE) are familiar to the community.

New literature continues to emerge that highlights the challenges and opportunities these fluctuating PE landscapes offer to the physical metallurgy of concentrated solid solutions [9,26,27]. Some notable progress has been made in recent works in treating variable PE landscape problems. For instance, Varvenne et al. [28] have developed an effective medium model, based on the embedded atom method (EAM) [29–31], which homogenizes the effects of solute in randomly arranged solid solutions. Several studies have leveraged this model to calculate material properties in concentrated systems including lattice parameters [16,32–35], elastic constants [16,32–35], cohesive energies [16,32], and stacking fault energies [16,34,35]. While this method is suitable for the calculation of homogenized bulk properties, a limitation arises when predicting material phenomena at length scales that approach those of the nanoscale fluctuations in the PE landscape. Certainly, there is a building body of literature focused on the study of material properties that are controlled by the fluctuations in the PE landscape of concentrated solid solutions. For example, Curtin and co-workers [6,8,13] have developed a method to measure the fluctuations in dislocation glide barriers that emerge from varied dislocation-solute and solute-solute interactions in solid solutions. Additionally, Zhang et al. [9] have incorporated a stochastic coefficient function into the classic Peierls-Nabarro model to predict a distribution of Peierls stresses in HEAs. Ghazeseidi and coworkers [36,37] report the statistics of a 'local' stacking fault energy that arises due to the varied solute interactions in the vicinity of planar defects. The qualitative link between these local stacking fault energies and the varied local chemical environment in concentrated solid solutions has been studied by Zhao et al. [17] and Ma and coworkers [38,39]. Undoubtedly, this new body of literature highlights the emergence of a variable 'local' property that arises from the statistical distribution of solute interactions that are inherent to concentrated solid solutions.

Although the observational connection between fluctuations in the PE landscape and the emergence of local material properties is now well established, there is a paucity of methods to directly calculate these fluctuations from system topology inputs. Specifically, analytical techniques that link the statistical variations in composition and neighbor pairings in solid solutions with fluctuations in PE landscapes remain sparse. To the authors' knowledge, the only example of a direct analytical relation is found in a recent study by Curtin and co-workers [6], where the specific use-case of fluctuations in solute-solute interactions during dislocation slip is studied. This dearth of analytical approaches limits the broad exploration of new theoretical models, where inputs for compositional and structural PE landscape fluctuations must be measured from atomistics [8] or take on assumed forms [9].

Here, we present an analytical approach to quantify the statistics of fluctuations in PE landscapes of randomly arranged solid solutions. This framework is developed here for EAM-type relations, which are commonly used to model the metallic bonding of alloys. These relations are applicable to randomly arranged solid solutions in any crystal system with arbitrary numbers of components and concentrations. To demonstrate this approach, we consider two PE landscapes: those of the cohesive energy and the GPFE. These landscapes are selected due to their roles in phase stability and in defining the intrinsic competition between deformation mechanisms. For the purposes of analyzing phase stability, examination of cohesive energy landscapes is performed on both face-centered cubic (FCC) and hexagonal close-packed (HCP) solid solutions. Application of this method to other PE landscapes is possible, with some limitations discussed later. The outcomes of this work provide analytical tools to study the statistics of PE fluctuations, which are important to the advancement of several classes of theories in the physical metallurgy of solid solutions.

## 2. METHODOLOGY

The primary challenge in quantifying the fluctuations in the PE landscape lies in the nanoscale variations in composition and neighbor pairings that are encountered when sampling small volumes in solid solution ensembles. Indeed, sampling of $n_c$ solute atoms in a randomly arranged solid solution produces a Gaussian distribution of mean compositions of the form $G^X \sim \left(\bar{C}^X, \sqrt{\frac{\bar{C}^X(1-\bar{C}^X)}{n_c}}\right)$, where $\bar{C}^X$ is the system-level composition of element $X$ presented as a molar fraction. In this notation, the first parameter is the distribution average, and the second parameter is the standard deviation. Furthermore, capitalized statistical parameters refer to population properties and lowercase symbols refer to samples drawn from these populations. By a similar calculation, the solute pairings from $n_s$ samples of two randomly selected solute pairs form a distribution with parameters $G^{XY} \sim \left(\bar{C}^X \bar{C}^Y, \sqrt{\frac{\bar{C}^X \bar{C}^Y(1-\bar{C}^X \bar{C}^Y)}{n_s}}\right)$. Figure 1 presents these statistics for a numerical simulation of a theoretical ternary system with randomly drawn samples, which mimics the organization of solute in a randomly arranged solid solution. As shown in the figure, the fluctuations in composition and solute pairings around the system averages are significant, which underscores the varied chemical environment encountered in solid solutions. A derivation of the statistical parameters (i.e., the standard deviation) of these distributions is provided in the Supplementary Materials.

Within the context of energy landscapes, the composite effect of the statistical variations in local composition and solute pairings give rise to the fluctuations observed in the PE. An example of this effect on the per atom binding energy ($E_a$) of a solute is provided in Figure 2. Here, the local composition and solute pairings within the first two nearest neighbor shells ($\zeta$) in a FCC lattice have been systematically varied for a binary FeNi system using the EAM potential of Bonny et al. [40]. In consideration of the statistical distributions shown in Figure 1, each of these local arrangements is possible in a randomly arranged solid solution, albeit with differing probabilities. As shown in Figure 2, the change in $E_a$ for a Ni atom varies with the sample composition ($\bar{c}^{Ni}$, primary x-axis of Figure 2), and with the partitioning of solute between the first ($\bar{c}_1^{Ni}$, secondary x-axis of Figure 2) and second nearest neighbor shells. The changes in $E_a$ can be significant - reaching as high as ~165 meV/atom in some configurations. It is this observation of the connection between solute arrangement and the binding energy that informs our approach to predicting the statistics of fluctuations.

### 2.1. Statistical analysis in the embedded atom method

The key to our approach to analyze PE fluctuations lies in recasting the equations of the interatomic potential in terms of the coordination environment around an atom such that varied compositions and neighbor pairings may be considered. To quantify these PE landscape fluctuations the statistical average and standard deviation are of interest. The inspiration for our approach follows from Varvenne et al. [28], where solute effects are treated through the creation of an EAM-based average-atom interatomic potential, which enables estimates of the homogenized properties of the bulk. Under the EAM, PE is defined by the summation of two terms: the embedding energy ($F$), which is a function of the charge density ($\rho$) from neighbor contributions; and the interaction energy ($V$) that arises from the pairwise interactions between atoms. In their usual presentation, the EAM terms are summed over all atoms and element types in a system to deliver the total system PE. For the purposes of this study, we are interested in determining the per atom PE using similar relations. To achieve a per atom representation, we reparametrize these relations through the introduction of the coordination number ($N_\zeta$), which provides the number of atoms in concentric shells at fixed distances from a central atom. These concentric shells of coordination are a superset of the nearest neighbor shells described above and are incremented with increasing radial distance. As demonstrated in the upcoming discussion, this relatively simple modification allows for the statistics of varied local solute environments to be analyzed. Moreover, this reparameterization can be leveraged to study a variety of PE landscapes, provided the coordination relations are known. This information is readily available for perfect crystals, which directly delivers the statistics of the cohesive energy landscape, and can be obtained for some crystal defects such as planar faults. This process is demonstrated in Section 2.2, where relations for the statistics of a varied GPFE landscape in an FCC system are provided. While this approach implicitly neglects lattice distortion effects on statistical parameters, results show that this assumption is reasonable for the systems studied herein.

The statistical analysis begins with an examination of the average per atom energy in a multicomponent system through summation of the modified EAM relations over all element types and coordination shells. Under the assumption of a random arrangement of solute, the average per atom binding energy ($\bar{E}_a$) can be calculated as:

$$\bar{E}_a = \sum_X \bar{C}^X F^X(\rho) + \frac{1}{2} \sum_{X,Y,\zeta} V_\zeta^{XY} N_\zeta \bar{C}^X \bar{C}^Y \qquad (1)$$

where $\bar{C}$ is the average concentration of a solute as defined previously, and $X$ and $Y$ are the indices of the different solute/element types. $V_\zeta^{XY}$ represents the pair interaction energies of solutes at distances corresponding to the various coordination shells. In this work, the summation of coordination shells is restricted by a cutoff distance of no less than $\sqrt{5/2}\, a_o$, where $a_o$ is the lattice parameter (here, $a_o \approx 0.35$ nm). This results in a summation of at least the first five coordination shells in all crystals considered, which was found to surpass any cutoff sensitivity. Following the assumption of Varvenne et al. [28], the per atom charge density term, $\rho$, is replaced with an average per atom charge density ($\bar{\rho}$). In most circumstances, this replacement simplifies the statistical analysis without a significant loss to model accuracy. The average per atom charge density follows as:

$$\bar{\rho} = \sum_{X,\zeta} \rho_\zeta^X \bar{C}^X N_\zeta \qquad (2)$$

where $\rho_\zeta^X$ represents the charge density contributed by solute $X$ in coordination shell $\zeta$. In Eqs. (1) and (2), the average concentrations offer the weighted probabilities of finding solute $X$ in coordination shell $\zeta$ for each term in the EAM relations.

Within the context of previous work, the relations provided in Eqs. (1) and (2) can be viewed as per atom analogs to the system-level homogenized relations developed in Varvenne et al. [28]. Analysis of the statistical scatter in per atom energies proceeds from separate calculations of the standard deviations of the embedding energies ($\Delta F$) and pair interaction energy terms ($\Delta V$). Under the constant charge density assumption, the standard deviation of the embedding energy is given as:

$$\Delta F = \sqrt{\sum_X (F^X - \bar{F})^2 \bar{C}^X} \tag{3}$$

where $F^X$ is the embedding energy of element $X$ evaluated at $\bar{\rho}$, and $\bar{F} = \sum_X F^X \bar{C}^X$ is the average per atom embedding energy. This expression for $\Delta F$ may be interpreted as a weighted combination of the variances of individual solute embedding energies. Due to the varied pair interactions between different solutes in coordination shells, analysis of the standard deviation of the pair interaction energy term requires a systematic approach that considers each pairing that contributes to statistical scatter. As demonstrated in Figure 1, the local concentration in the vicinity of a solute atom may vary significantly from the global chemistry, and the partitioning of solute into differing coordination shells (see Figure 2) presents additional pathways for statistical scatter. These added considerations underscore the complexities that arise in determining the PE statistics of solid solutions. The final relation for the standard deviation of the pair interaction energy is given as:

$$\Delta V = \sqrt{\sum_X \bar{C}^X (\Delta V^X)^2 + \sum_{X<Y} \bar{C}^X \bar{C}^Y (\bar{V}^X - \bar{V}^Y)^2} \tag{4}$$

where $\bar{V}^X$ and $\Delta V^X$ are the average and standard deviations of the per atom interaction energies, respectively, when solute $X$ is the central atom. A detailed derivation of these terms is provided in Appendix A. Using the standard rules for the combinations of statistical variables, Eqs. (3) and (4) may be combined to provide the standard deviation of the per atom energy

($\Delta E_a$):

$$\Delta E_a = \sqrt{\Delta F^2 + \left(\frac{\Delta V}{2}\right)^2 + 2cov\left(F, \frac{1}{2}V\right)} \quad (5)$$

where $cov\left(F, \frac{1}{2}V\right)$ is the covariance between the embedding and the pair interaction energies. The $1/2$ coefficient appears in the $\Delta V$ and covariance terms to avoid double counting of interaction energies, as in Eq. (1). The covariance term can be calculated directly from the overall composition and the interatomic potential coefficients as:

$$cov\left(F, \frac{1}{2}V\right) = \frac{1}{2}\sum_X \bar{C}^X F^X \bar{V}^X - \left(\sum_X \bar{C}^X F^X\right)\left(\frac{1}{2}\sum_{X,Y,\zeta} V_\zeta^{XY} N_\zeta \bar{C}^X \bar{C}^Y\right) \quad (6)$$

A full derivation of this expression is provided in Appendix B.

Collectively, Eqs. (1) - (6) provide a method to evaluate the statistical variations of per atom potential energies in randomly arranged solid solutions. These relations are derived directly from the EAM formulation and require only coefficients from the interatomic potential, solute concentrations, and coordination relations as inputs. Furthermore, there are no adjustable parameters or new assumptions introduced to produce these statistical relations. These equations are general and can be applied to any crystalline structure, provided the coordination relations are given. Indeed, the application of these relations to planar faults is a necessary step in the analysis of the GPFE landscape that follows in the next section.

**2.2. Extension of modified EAM relations to the GPFE landscape**

The PE landscapes considered in this study are those of the cohesive energy and the GPFE. The cohesive energy landscape is a composite of per atom binding energy values that describe the potential energy change due to deposition from the gas phase into a crystalline phase, which serves as a reference for comparison of phase stability. The statistical parameters of the cohesive energy landscape can be therefore retrieved directly from Eqs. (1) - (6). The GPFE landscape is formed from the area normalized interplanar excess energy (i.e., the planar fault energy, $\gamma$) arising from the coordinated shearing of the pristine FCC lattice by Shockley partial dislocations along <112>/{111}, as required to form and thicken deformation twins [41,42]. Within the context of deformation mechanisms, the GPFE landscape can be used to examine the intrinsic competition between dislocation slip and deformation twinning in a FCC material, which has important implications on the evolution of mesoscale plasticity in low stacking fault energy systems. The tendency for a material to exhibit deformation twinning is described through a variety of twinnability parameters [43–48], which are dependent on the critical energies (i.e., peaks and valleys) of the GPFE landscape. Figure 3 provides a schematic of a typical GPFE landscape for an FCC material, with the critical energies labeled. These energies are the unstable stacking fault energies ($\gamma_{usf}^{0.5}$ and $\gamma_{usf}^{1.5}$), the intrinsic and extrinsic stacking fault energies ($\gamma_{isf}^{1}$ and $\gamma_{esf}^{2}$), and the unstable twinning and twinning fault energies ($\gamma_{utf}^{2.5}$ and $\gamma_{tf}^{3}$). Here, the superscript refers to the crystallographic slip that is required to create each fault, referenced as a multiple of the Shockley partial Burger's vector ($b_{112}$). Fluctuations in these critical energies are of interest to the community, as they have implications on deformation mechanism competition [43–46,48], the waviness of stacking faults ribbons [8,36], variations in twinning stresses [49], and the local stability of FCC and hexagonal close-packed (HCP) phases [50,51], amongst other elements of mechanical metallurgy.

Analysis of the fluctuations in the critical energies of the GPFE begins with the recognition that the configuration of solute around each of the associated planar faults can be fully described by sets of coordination relations. Therefore, the statistical relations of the previous section can be leveraged to analyze fluctuations in the inflection points of the GPFE landscape. However, in contrast to the pristine FCC lattice, where relations are spatially invariant, the coordination environment changes for atoms lying in each of the {111} crystallographic planes near a planar fault due to changes in the local stacking sequence. Figure 4 illustrates the changes in the coordination relations in the {111} layers near an unstable stacking fault. Here, we use a two-index notation ($n$, $m$) to refer to faulted planes, where $n$ refers to the number of Shockley partial dislocations required to create the defect, and $m$ is a {111} layer enumerated sequentially from the shearing plane ($m = 0$). As shown in Figure 4, the disruption of the normal stacking sequence gives rise to new peaks in the coordination relations. Intuitively, the appearance of non-FCC peaks occurs at larger radial distances in layers further from the shearing plane, until they are pushed outside of the cutoff radius $\left(\sqrt{5/2}\, a_o\right)$ at 3 {111} layers (i.e., a distance of $\sqrt{3} a_o$ for undistorted/idealized crystal). In classifying FCC and non-FCC coordinated crystallographic layers, a comparison of coordination relations within the cutoff radius is used as the guiding criterion. Other schemes (e.g., based on stacking sequence) may lead to different classifications. Coordination relations for the other interplanar defects associated with the critical energies of the GPFE landscape (i.e., the unstable stacking and twinning faults, extrinsic stacking fault, and twin fault) are provided in the Supplementary Materials.

Calculation of the average planar fault energy ($\bar{\gamma}$) follows directly from a comparison of average per atom energies in the faulted and fault-free conditions as:

$$\bar{\gamma}^n = \rho_{111} \sum_m (\bar{E}_a^{n,m} - \bar{E}_a^{FCC}) \tag{7}$$

where $\bar{E}_a^{FCC}$ and $\bar{E}_a^{n,m}$ are the average per atom energies for an FCC coordination and the coordination in the $m^{th}$ {111} layer near a $n$-type fault, respectively. $\rho_{111}$ is the number density of atoms in the FCC {111} plane, which is given as $4/\sqrt{3}a_o^2$. $\bar{E}_a^{FCC}$ can be calculated directly from Eq. (1). Determination of $\bar{E}_a^{n,m}$ follows from a separate application of Eq. (1) to each of the coordination environments in the $m$ layers around a planar defect (see Figure 4), with $F^X = F^{X,n,m}$, $V_\zeta^{XY} = V_\zeta^{XY,n,m}$, and $N_\zeta = N_\zeta^{n,m}$.

Calculation of the standard deviation of the planar fault energy ($\Delta\gamma$) begins with a determination of the standard deviations of per atom energies in each of the $m$ {111} crystallographic layers influenced by a $n$-type fault (i.e., $\Delta E_a^{n,m}$). For standard deviation calculations, we consider only those {111} layers whose coordination relations differ from the pristine FCC case within the specified cutoff radius to avoid the convolution of cohesive energy and excess energy distributions. The $\Delta E_a^{n,m}$ terms can be retrieved through the application of Eqs. (1)-(6) in Section 2.1 using the same modifications to the embedding and pair interaction energies and coordination relations as in the calculation of $\bar{E}_a^{n,m}$. Using the same rationale for calculating the aggregated statistics of non-overlapping sub-populations in per atom interaction energies (see Eq. (4) and Appendix A), the standard deviation of per atom energies ($\Delta E_a^n$) over all non-FCC layers in the vicinity of a planar fault is calculated as:

$$\Delta E_a^n = \sqrt{\frac{1}{M}\left[\sum_m (\Delta E_a^{n,m})^2 + \frac{1}{M}\sum_{m1<m2} (\bar{E}_a^{n,m1} - \bar{E}_a^{n,m2})^2\right]} \tag{8}$$

where $M$ is the total number of non-FCC coordinated layers created by the planar fault.

Following the rules for the addition and subtraction of statistical variables, we arrive at a relation for the standard deviation of the planar fault energy ($\Delta\gamma$), which can be calculated from the sum of variances in the faulted and fault-free conditions:

$$\Delta\gamma^n = \sqrt{(\Delta E_a^n)^2 + (\Delta E_a^{FCC})^2 - 2cov(E_a^n, E_a^{FCC})}\, \rho_{111} \qquad (9)$$

where $cov(E_a^n, E_a^{FCC})$ is the covariance between faulted and non-faulted per atom energy distributions. The covariance between these distributions can be determined from the difference between the expectation value of products and the product of expectation values. We find a good approximation to this exact definition from the following relation:

$$cov(E_a^n, E_a^{FCC}) \approx \frac{1}{M}\left(\sum_{X,m} \bar{C}^X \bar{E}_a^{X,n,m} \bar{E}_a^{X,FCC} - \bar{E}_a^{FCC} \sum_m \bar{E}_a^{n,m}\right) \qquad (10)$$

where $\bar{E}_a^{X,n,m}$ and $\bar{E}_a^{X,FCC}$ are the solute-level average per atom energies in the faulted and FCC-coordinated conditions. A detailed derivation of this relation is provided in Appendix B.

The statistical analysis presented in Sections 2.1 and 2.2 provide a method to calculate the statistics of fluctuations in the cohesive energy and GPFE landscapes. These landscapes provide insights into the phase stability and deformation mechanism competition in randomly arranged solid solutions. These relations are derived for EAM-type interatomic potentials, which are relevant for metals and alloys, and can be applied to systems of arbitrary components and compositions. Furthermore, our approach is general, and can be expanded to consider other PE landscapes, provided the system topology can be assembled from sets of coordination relations. Although calculations are somewhat tedious, these statistical relations can be easily scripted into algorithms for computation. Python-based algorithms to calculate the statistical parameters of cohesive energy and GPFE landscapes for randomly arranged solid solutions will be made available with the published version of this manuscript. In subsequent sections, results and validation datasets are presented for the cohesive energy and GPFE landscapes in the NiCo and FeNiCr systems.

**2.3. Validation of the statistical relations**

The statistical relations developed in previous sections are validated using molecular statics (MS) simulations with the Large-scale Atomic/Molecular Massively Parallel Simulator (LAMMPS) software package [52]. Visualization of simulation topologies is realized using The Open Visualization Tool (OVITO) [53]. Two simulations are performed: simple MS relaxation calculations and interplanar shearing simulations, which are used to validate predictions of fluctuations in the cohesive energy and GPFE landscapes, respectively. The procedures for these simulations are well established within the community (e.g., Refs. [2,14–16,28,34]). System sizes are chosen based on convergence studies (see Supplementary Materials). Data presented in the main text corresponds to systems measuring at least 10 by 10 by 5 nm (> 50,000 atoms) and 35 by 35 by 7 nm (> 800,000 atoms) in size for MS relaxation and interplanar shearing simulations, respectively. MS relaxation calculations are performed using the conjugate gradient minimization tool built within LAMMPS. Simulation boundaries are periodic in all directions and are allowed to expand/contract to minimize stresses. For interplanar shearing simulations, we have followed the same methodology from our previous work [47]. Briefly, single-crystal FCC systems are constructed such that the <110>, <112>, and <111> crystallographic directions are aligned to the global $x$, $y$, and $z$ axes, respectively. Periodic boundary conditions are enforced on the $x$ and $y$ boundaries and the $z$ surfaces are free surfaces. The GPFE curve is obtained through a three-stage rigid shearing process, where the shear per stage is equal to the magnitude of a <112>-type Shockley partial dislocation (*i.e.*, $\frac{a_o}{\sqrt{6}}$). This shearing process begins its first stage by displacing two halves of the crystal along the <112> direction within a {111} shear plane. The second and third stages are completed by shearing the crystal at the plane immediately adjacent to the fault formed in the prior stage, as required to realize a twin fault. Each shearing stage is executed over 100 shearing increments to capture the intermediate configurations and excess energies of the GPFE landscape. In between each shearing step, the system is relaxed in the *z*-direction and the per atom potential energies are recorded. The GPFE is then obtained from the per atom energies of the relaxed configurations (i.e., the distribution $E_a$). The average planar fault energy (i.e., $\bar{\gamma}$) can be directly calculated over the faulted layers using the relation in Eq. (7). In MS simulations, the standard deviation of planar fault energies (i.e., $\Delta\gamma$) is determined from its definition as the expectation value of deviations from the mean: $\Delta\gamma = \sqrt{\langle([E_a^n - \bar{E}_a^n] - [E_a^{FCC} - \bar{E}_a^{FCC}])^2\rangle}\rho_{111}$, where $\langle\rangle$ refers to the expectation value operator. The standard deviation is calculated here from specific {111} layers using the same cutoff criteria as described in Section 2.2.

## 3. RESULTS AND DISCUSSION

For the purposes of this study, we have selected the binary NiCo and ternary FeNiCr systems as our benchmark materials. One key selection criteria for these systems is their relevance in deformation twin [54–56] and stress-induced HCP martensite forming materials [56–59], for which the fluctuations in the fault energies of the GPFE landscape are of inherent interest. In addition to these characteristics, the FeNiCr system is selected due to its technological importance in stainless steels and the availability of computational datasets from which results may be validated [28,32,33,60,61]. The NiCo system also offers the opportunity to study the accuracy of our method in polymorphic solid solutions, with a FCC to HCP transition known in this system at ~ 70 at.% Co [62]. The following sections present the application of the equations of Sections 2.1 and 2.2 to the FeNiCr and NiCo systems over a wide compositional range. Although the FeNiCr system exhibits complicated multiphase microstructures at some compositions, our analysis focuses on the FCC and HCP structures as a theoretical exercise. Validation datasets have been generated from MS relaxation studies using the procedures of Section 2.3. For these simulations, the EAM-based potentials of Bonny et al. [40] (FeNiCr) and Béland et al. [63] (NiCo) are used to model interatomic interactions.

### 3.1. Statistical analysis of the cohesive energy landscape

The statistical parameters of the cohesive energy landscape are given directly by the average and standard deviations of per atom binding energies (i.e., $\bar{E}_a$ and $\Delta E_a$). Figure 5 provides these quantities for the FCC phase of the FeNiCr, and the FCC and HCP phases of NiCo. The FeNiCr compositional space is charted along a search path of Fe$_{(1-x)/2}$Ni$_{(1-x)/2}$Cr$_x$, with literature data provided from Ref. [28] for validation purposes. Results are obtained for NiCo across its entire compositional space. Using the statistical relations of Section 2.1, analytical calculations are displayed alongside the results of MS relaxation simulations. As shown in the figure, $\bar{E}_a$ is shown to vary monotonically from ~ -4.5 to -3.7 eV/atom for the FeNiCr data. The NiCo system exhibits ranges of ~ -4.45 to -4.38 and ~ -4.43 to -4.39 eV/atom for the FCC and HCP phases, respectively. MS relaxation data show excellent agreement with the predictions of the statistical relations for both systems. Indeed, the analytical estimates of per atom average energies deviate by less than 0.5% for both FeNiCr and NiCo. Furthermore, the statistical relations accurately capture the transition in FCC to HCP phase stability at Co-rich compositions of > 70 at.%.

Standard deviations in the per atom energies were found to vary between 0 to ~0.46 eV/atom and 0 to ~ 0.043 eV/atom across the sampled FeNiCr and NiCo systems, respectively. The large standard deviations in FeNiCr datasets are driven by the dramatic change in per atom energies with increasing Cr content. Conversely, NiCo per atom energies vary only slightly across its compositional space, which leads to lower standard deviations. As with the average per atom energies, the standard deviations in the MS data show excellent agreement with analytical calculations. Indeed, deviations in analytical estimates of standard deviations are < 5% for both FeNiCr and NiCo, except at singularities near the terminal compositions where the standard deviations approach zero. Comparison between datasets shows that these analytical relations accurately capture the order of magnitude changes in statistical fluctuations across different systems and chemistries. Some features of this agreement bear noting. Namely, the statistical relations accurately predict a vanishing standard deviation at pure chemistries (i.e., right of FeNiCr plot and terminuses of the NiCo plot). Furthermore, our statistical relations also capture the asymmetry in the standard deviation of the NiCo system. More broadly, the strong agreement between our idealized analytical approach and relaxed MS datasets demonstrates the importance of the distribution of solute arrangements in driving PE fluctuations. Moreover, these results also indicate that lattice distortion effects hold a minor role in the determination of per atom energies. One interpretation is that although lattice distortion certainly exists in these solid solutions, the energetic fluctuations induced by disruptions to the lattice on an atom-by-atom basis are self-reconciling. However, caution must be taken when generalizing this finding to other PE landscapes or systems with more pronounced mismatches in atoms size, where lattice distortion effects may exhibit a stronger effect on the emergence of relaxed energies.

In addition to insights into lattice distortion effects, these results provide new tools to quantify phase stability, which contributes to the ongoing dialogue of metastable microstructures in concentrated solid solutions. For instance, recent computational studies of cohesive energies in equimolar CrCoNi have provided some conflicting results. While the HCP phase of this system appears to be the preferred low-temperature equilibrium structure [64], the high-temperature FCC phase is typically observed to be dominant in experiments [65]. One interpretation is that the narrow differences in HCP/FCC energies enable the retention of a metastable FCC phase that is realized upon cooling from high-temperature materials processing. The results of this section add new quantitative tools to analyze this metastability argument. For instance, as demonstrated in the NiCo system, local variations in chemistry and solute pairings create sufficiently large statistical fluctuations in PE such that the FCC and HCP phases find local regions of favorable/unfavorable stability across a system topology.

The success of our statistical relations in replicating the per atom energies of MS relaxation calculations motivates a computational search of compositional spaces without the need for exhaustive atomistic simulations. Figure 6 presents the statistical parameters of the cohesive energy landscape in ternary plot form for the entire compositional space of the FeNiCr system, with the compositions reported in Figure 5 traced with a dashed arrow. Some interesting features emerge from this complete search of the FeNiCr system. Namely, the largest standard deviations in the per atom energies are found along the binary NiCr terminus of this ternary diagram. This result may be understood from an examination of the per atom binding energies in the FeNiCr system, which reach their minimum and maximum values at the pure Ni (~ -4.5 eV/atom) and Cr (~ -3.7 eV/atom) compositions, respectively. Consequently, to a first approximation, the standard deviation can be expected to reach a maximum at an equimolar chemistry of NiCr. We believe that this finding is consequential within the broader context of PE landscapes in solid solutions. For instance, recent theories have demonstrated the role of large PE fluctuations in driving the unusual strengthening of HEAs. Examples include the solute-dislocation interaction model of Varvenne et al. [8,13] and the stochastic Peierls-based model of Zhang et al. [9]. The success of these theories, combined with the community's enthusiasm for HEAs, has led to perhaps an overly restrictive perspective on opportunities for property improvements. That is, increasing the number of components best realizes the benefits of solid solutions by maximizing solute dispersion. With some caveats, we offer this data as a theoretical counter example, where a binary sub-alloy exhibits larger PE fluctuations than a parent ternary. Collectively, these findings show that solute-induced PE fluctuations represent a complex interplay of pair interactions, solute arrangement, and ensemble chemistry.

**3.2. Statistical parameters of planar fault energies in FCC solid solutions**

Following the procedures outlined in Section 2.2, the standard deviations in the critical planar fault energies of the GPFE landscape have been studied for four different FCC solid solutions. The selected system chemistries are $Fe_{73}Ni_8Cr_{19}$, and equimolar FeNiCr, NiCr, and NiCo. The inclusion of NiCr is motivated by the peak statistical fluctuations observed at this chemistry in the cohesive energies of the previous section. $Fe_{73}Ni_8Cr_{19}$ has been selected due to its technological relevance in 304-type stainless steels, where deformation twinning is a competing plasticity mechanism. As the first step in the determination of fault energies, Figure 7 provides the standard deviations of per atom energies at various faulted conditions for each of these systems. The data from MS shearing simulations is overlaid with predictions from statistical relations, showing excellent agreement. The largest deviations between datasets are found in the equimolar NiCo, but do not exceed 7% error, with most deviations being < 2%. Standard deviations range from between ~ 0.04 to 0.55 eV / atom, with the lower and upper extremes found in the equimolar NiCo and NiCr binary systems, respectively. These findings align with the cohesive energy data of pristine FCC systems, which showed similar ranges. Indeed, the introduction of planar faults was found to only change per atom energies by a maximum of ~ 2 % in the FeNiCr chemistries and ~ 6% in equimolar NiCo. However, it should be noted that fluctuations in individual layers around a *n*-type fault may hold larger scatter than the aggregated variance of all *M* layers within the cutoff distance of the defect, from which $\Delta E_a^n$ is reported.

Building upon the calculations of $\Delta E_a^n$, we report the statistical scatter in the critical fault energies of the GPFE landscapes ($\Delta \gamma^n$) in Figure 8 for each system. The analytical calculations of $\Delta \gamma^n$ are shown as errorbars and are overlaid with the GPFE landscapes calculated from interplanar shearing simulations. Literature values for the average intrinsic stacking fault energies in each system have been provided for comparison purposes. Some discontinuities in the scatter of the GPFE landscape appeared at the stable planar faults, which we associate with discontinuities in the numbers of non-FCC coordinated layers upon transitions through layer-by-layer shearing operations. Although the analytical estimates of standard deviations exhibit only a modest agreement with the MS data, we find this result to be reasonable as the source of error is not related to any new failings in our theory. Rather, these errors arise due to the nature of the standard deviation calculation in Eq. (9) and its extreme sensitivity to small changes in input parameters. Indeed, while analytical predictions for per atom energies and covariances (see Figure 7 and Figure B1) are in excellent agreement with MS relaxation calculations, the subtractive elements of Eq. (9) greatly amplify any small discrepancies. The units of these fluctuations also provide useful context for consideration. That is, predictions of statistical scatter on the order of mJ over areas of $m^2$ would seem to be difficult to determine with great precision.

Despite only a modest numerical agreement between datasets, the trends and broader context of these results merit some discussion. In both approaches, the sequencing of the magnitudes of GPFE fluctuations are in agreement. Furthermore, both approaches predict that the NiCr system, which showed the highest fluctuations in cohesive energies, exhibits the lowest fluctuations in critical fault energies for each of the FeNiCr chemistries considered. This reduction in scatter is a result of the strong correlations observed in the NiCr system (see Figure B1). An interesting interpretation of this result is that strongly correlated systems with large fluctuations in per atom binding energies do not exhibit proportionally large fluctuations in their planar fault energies. An additional point of discussion from these results is the observations of fluctuations causing negative stacking fault energies in both the analytical calculations and MS data. This is particularly relevant for the $Fe_{73}Ni_8Cr_{19}$ measurements as negative stacking fault energies are linked to the well-known stress-induced HCP martensite transformation exhibited by related stainless steel chemistries [56–59].

Within the broader context of concentrated solid solutions, we believe that the statistical relations developed in this work contribute to an ongoing discussion within the community on the interpretation of planar fault energies. For instance, recent studies of planar fault energies in medium- and high entropy alloys have reported a wide scatter in values [18,50,64,66]. Discussions in the newest literature have brought some clarity [36–39]. Namely, planar fault energies are sensitive to the arrangement of solute atoms near the fault plane, which gives rise to the concept of a "local" planar fault energy. Our results here offer a method by which the statistics of these local planar fault energies can be analytically studied. This contribution finds direct applications in quantifying the unusual metallurgy of concentrated solid solutions, with examples including waviness in stacking fault ribbons [8,36], and variations in solute-fault interaction energies [49].

## 4. CONCLUSIONS

An analytical approach has been developed to quantify the statistics of PE landscapes in randomly arranged solid solutions. For this purpose, the equations of EAM-based interatomic potentials are reparametrized to yield a set of relations defined by the coordination environment of solute. From these recast equations, exact relations for the statistical parameters of the embedding and pair interaction energy are derived. This set of initial relations is general and can be applied to study the cohesive energy landscapes of any crystalline system with an arbitrary number of components and composition. Using coordination as a key input parameter, we expanded the applicability of statistical relations to provide an analytical method to examine fluctuations in the excess energies of planar faults. Collectively, these analytical tools provide the community a method to rapidly quantify PE fluctuations in pristine and defective solid solution systems described by sets of coordination relations, which finds applications in studies of phase stability, and competition between deformation mechanisms.

In the second portion of this study, these relations are leveraged to study the cohesive energy and GPFE landscapes of FeNiCr and NiCo. In most cases, predictions from statistical relations were in excellent agreement with validation datasets that were collected from MS relaxation and interplanar shearing simulations. From this effort, some key insights into the metallurgy of solid solutions have emerged that contribute to ongoing discussions within the community. For instance, our results for NiCo offer a nuanced perspective on phase stability. That is, local variations in composition and solute pairings create sufficiently large fluctuations in cohesive energy that competing lattices find regions of favorable/unfavorable stability across a system topology. Within the context of the role of PE fluctuations in the physical properties of solid solutions, results from a theoretical search of the FeNiCr composition space contribute to a building theme in the literature. Namely, maximizing the effects of solid solution ensembles is more nuanced than simply selecting chemistries with the greatest solute dispersion. An additional outcome of this effort is a method to quantify the statistics of local planar fault energies, which is a new concept that has emerged from the varied stacking fault energies reported in the recent HEA literature.

**ACKNOWLEDGEMENTS**

This material is based upon work supported by the National Science Foundation under Grant No. DMR-2144451. The authors gratefully acknowledge the Advanced Cyberinfrastructure for Education and Research (ACER) group at the University of Illinois at Chicago for providing computational resources and services needed to deliver research results delivered within this paper. URL: https://acer.uic.edu. The authors would also like to thank Ariana Sofia Del Valle for her assistance in maintaining and updating python scripts.

**AUTHOR CONTRIBUTIONS**

This study was conceived by M. D and was executed by R. J. and C. W. The statistical analysis and atomistic simulations were performed by R. J. and C. W. under the direction of M. D. The manuscript was written by R. J. and M. D. with contributions from C. W.

**ADDITIONAL INFORMATION**

Supplementary information is available in the online version of the paper.

**COMPETING INTERESTS**

The authors declare no competing interests.

**APPENDIX A: STANDARD DEVIATION OF THE PER ATOM PAIR INTERACTION ENERGY**

The calculation of the standard deviation of the pair interaction energy term $(\Delta V)$ begins with a consideration of the types of pairwise interactions that are available to specific solutes in randomly arranged solid solutions. For instance, a solute of type $X$ may interact with other solutes of type $X$ or $Y$, forming pair interactions $XX$ or $XY$. However, through the specification of atom type $X$ as the solute of consideration, interactions of $YX$ or $YY$ are not considered and belong to a separate set of solute pairings (i.e., as in conditional probabilities). This nuanced point is significant, as it highlights the nature of the statistical scatter that arises in pair interactions. That is, the statistical scatter of interactions can be viewed as an aggregation of separate solute pairing distributions that can be individually specified. Phrased differently, calculation of $\Delta V$ can be achieved through a separate determination of $\Delta V^X$, where $\Delta V^X$ is the standard deviation of interaction energies between a central solute $X$ and all other solute neighbors within the coordination cutoff. One additional complication is that the pair interaction energy is formed from the contributions of solute pairings from all neighbors. The distribution of interaction energies is therefore comprised of additive combinations, which can be treated using the central limit theorem of sums. Under these considerations, the solute-level standard deviations can be calculated through a weighted combination of variances in each of the coordination shells as:

$$\Delta V^X = \sqrt{\sum_{\zeta,Y}\left(V_\zeta^{XY} - \bar{V}_\zeta^X\right)^2 \bar{C}^Y N_\zeta} \tag{A1}$$

where $\bar{V}_\zeta^X = \sum_Y V_\zeta^{XY} \bar{C}^Y$ is the average interaction energy for a solute $X$ in coordination shell $\zeta$. In addition to the solute-level standard deviation, the solute-level average of the interaction energy $(\bar{V}^X)$ is also of interest. This term can be calculated from $\bar{V}_\zeta^X$ as:

$$\bar{V}^X = \sum_\zeta \bar{V}_\zeta^X N_\zeta = \sum_{\zeta,Y} V_\zeta^{XY} \bar{C}^Y N_\zeta \tag{A2}$$

In the formulation of Eq. (A1), it is assumed that the variances between coordination shells are uncorrelated (i.e., their covariance is zero), which is appropriate for randomly arranged solid solutions. In both equations, the transformation of the statistical parameters of individual solute pair energies to the distribution of additive combinations of per atom pair interaction energies is achieved through the coordination number term, $N_\zeta$, as per the central limit theorem of sums. Eq. (A1) provides the standard deviations of the distributions pair interaction energies for each solute, which can be considered as non-overlapping sub-populations of a larger distribution. Under this consideration, the aggregated standard deviation ($\Delta V$) of the system-level pair interaction energy term can then be calculated as a combination of the solute-level standard deviations as:

$$\Delta V = \sqrt{\sum_X \left[\Delta V^{X^2} + (\bar{V}^X - \bar{V})^2\right] \bar{C}^X} \qquad (A3)$$

where $\bar{V}$ is the system-level average pair interaction energy over all coordination shells. Eq. (A3) can be viewed as a weighted linear combination of individual solute-level variances. The second term (difference of averages, $\bar{V}^X - \bar{V}$) is a correction term to account for differences between the sub-population and aggregate averages. Eq. (A3) can be simplified further, without any loss of numerical accuracy, to eliminate the need to calculate $\bar{V}$. This simplified relation appears as Eq. (4) in Section 2.1.

**APPENDIX B: COVARIANCES OF STATISTICAL VARIABLES**

This appendix provides derivations for two covariance terms used in the paper body: $cov\left(F, \frac{1}{2}V\right)$ and $cov(E^n, E^{FCC})$. In both derivations, the covariance is derived from its definition as the difference between the expectation value of products and the product of expectation values of the statistical variables. Beginning with the embedding and interaction energy distributions (i.e., $cov\left(F, \frac{1}{2}V\right)$), the covariance can be calculated from its expectation value definition as:

$$cov\left(F, \frac{1}{2}V\right) = \langle \frac{FV}{2} \rangle - \langle F \rangle \langle \frac{V}{2} \rangle \tag{B1}$$

where the $\langle \rangle$ operator refers to the expectation value. Here, the individual expectation values of the embedding (i.e., $\langle F \rangle$) and pair interaction energies (i.e., $\langle V/2 \rangle$) are given directly by the terms of Eq. (1). The expectation value of products (i.e., $\langle FV/2 \rangle$) can be calculated from the sum of the products of $(F, V/2)$ pairings weighted by their probability. In these pairings, $F$ can be separated into solute-specific constants, whose probabilities are weighted by composition (i.e., $\bar{C}^X F^X$), as follows from the average per atom charge density assumption. Similarly, the pair interaction energy can also be decomposed into the separate distributions of solute-level interaction energies (i.e., $V^X$). These solute-level distributions are paired with $\bar{C}^X F^X$ coefficients, such that $\langle FV/2 \rangle = \frac{1}{2}\sum_{X,i} \bar{C}^X F^X p^i V^X$, where $p^i$ is the probability of $V^X$. The probability-weighted sum of $V^X$ can be calculated from its expectation value, $\langle V^X \rangle = \sum_i p^i V^X = \bar{V}^X$, which is given by Eq. (A2). Under these considerations, the summation of the products of these terms (i.e., $\frac{1}{2}\sum_X \bar{C}^X F^X \bar{V}^X$) returns the expectation value of products. The covariance can therefore be calculated as:

$$cov\left(F, \frac{1}{2}V\right) = \langle \frac{FV}{2} \rangle - \langle F \rangle \langle \frac{V}{2} \rangle \tag{B2a}$$

$$= \frac{1}{2}\sum_X \bar{C}^X F^X \bar{V}^X - \left(\sum_X \bar{C}^X F^X\right)\left(\frac{1}{2}\sum_{X,Y,\zeta} V_\zeta^{XY} N_\zeta \bar{C}^X \bar{C}^Y\right) \tag{B2b}$$

Analysis of the covariance between the distributions of faulted and FCC-coordinated per atom energies begins with its definition as the differences of expectation values:

$$cov(E_a^n, E_a^{FCC}) = \langle E_a^n\, E_a^{FCC} \rangle - \langle E_a^n \rangle \langle E_a^{FCC} \rangle \tag{B3}$$

Individual expectation values (i.e., $\langle E_a^{FCC} \rangle$ and $\langle E_a^n \rangle$) are given directly by Eq. (1). However, in the latter calculation, the average energies in each of the $m$ {111} layers influenced by the defect must first be calculated (i.e., $\bar{E}_a^{n,m}$). The average over all non-FCC layers (i.e., $\bar{E}_a^n$) is then calculated by the weighted sum of per atom average energies (i.e., $\frac{1}{M}\sum_m \bar{E}_a^{n,m}$).

The expectation value of products (i.e., $\langle E_a^n E_a^{FCC} \rangle$) can be separated into a summation of expectation values of the embedding (i.e., $F^n$, $F^{FCC}$) and pair interaction energy terms (i.e., $V^n$, $V^{FCC}$) in the faulted and fault-free conditions:

$$\langle E_a^n\, E_a^{FCC} \rangle = \left\langle \left(F^n + \frac{1}{2}V^n\right)\left(F^{FCC} + \frac{1}{2}V^{FCC}\right) \right\rangle \tag{B4a}$$

$$= \langle F^n F^{FCC} \rangle + \langle F^{FCC}\tfrac{1}{2}V^n \rangle + \langle F^n \tfrac{1}{2}V^{FCC} \rangle + \langle \tfrac{1}{2}V^n \tfrac{1}{2}V^{FCC} \rangle \tag{B4b}$$

Each of these expectation values of products can be further analyzed using arguments similar to those presented for the determination of $cov\left(F, \frac{1}{2}V\right)$ in Eq. (B2b). For instance, the expectation value of the product of embedding energy distributions (i.e., $\langle F^n F^{FCC} \rangle$) follows directly from the separation of these terms into their solute-level, and fault plane-level constants (i.e., $F^{X,n,m}$ and $F^{X,FCC}$), whose probability of pairing is $\bar{C}^X/M$. Here, $M$ is the total number of non-FCC coordinated {111} planes within the cutoff radius of the planar defect. The expectation value then becomes the summation of paired embedding energies over all solute and faulted layers, $\langle F^n F^{FCC} \rangle = \frac{1}{M}\sum_{X,m} \bar{C}^X F^{X,n,m} F^{X,FCC}$. The expectation values of cross-products (i.e., $\langle F^{FCC} V^n/_2 \rangle$ and $\langle F^n V^{FCC}/_2 \rangle$) follow from the expressions in Eq. (B2b), with the embedding and pair interaction coefficients updated to reflect the summation over the fault environment: $\langle F^{FCC} V^n/_2 \rangle = \frac{1}{M}\sum_{X,m} \frac{1}{2} \bar{C}^X F^{X,FCC} \bar{V}^{X,n,m}$ and $\langle F^n V^{FCC}/_2 \rangle = \frac{1}{M}\sum_{X,m} \frac{1}{2} \bar{C}^X F^{X,n,m} \bar{V}^{X,FCC}$. Here, the averages of pair interaction energies (i.e., $\bar{V}^{X,FCC}$ and $\bar{V}^{X,n,m}$) are given by application of Eq. (A2). For the final expectation of products term, $\langle V^n/_2 V^{FCC}/_2 \rangle$, we assume that these distributions are independent of one another (i.e., $cov(V^n/_2, V^{FCC}/_2) = 0$). This assumption enables a separation of the distribution products and a direct estimate for this term as $\langle V^n/_2 V^{FCC}/_2 \rangle \approx \langle V^n/_2 \rangle \langle V^{FCC}/_2 \rangle = \frac{1}{M}\sum_{X,m} \frac{1}{4} \bar{V}^{X,n,m} \bar{V}^{X,FCC}$. Although the distributions of $V^n$ and $V^{FCC}$ are likely correlated due to sharing some common neighbor pairings, we find that this approximation provides excellent estimates of covariances (see Figure B1). Combining each of these expectation values delivers an approximate solution for the expectation value of products of distributions of per atom energies. Without further approximation, this equation can be factored and simplified in terms of the solute-level average per atom energies of the faulted and FCC distributions (i.e., $\bar{E}_a^{X,n,m}$ and $\bar{E}_a^{X,FCC}$) as:

$$\langle E_a^n E_a^{FCC} \rangle \approx \frac{1}{M} \sum_{X,m} \bar{C}^X \left( F^{X,n,m} F^{X,FCC} + \frac{1}{2} F^{X,FCC} \bar{V}^{X,n,m} + \frac{1}{2} F^{X,n,m} \bar{V}^{X,FCC} \right.$$

$$\left. + \frac{1}{4} \bar{V}^{X,n,m} \bar{V}^{X,FCC} \right) \tag{B5a}$$

$$= \frac{1}{M} \sum_{X,m} \bar{C}^X \bar{E}_a^{X,n,m} \bar{E}_a^{X,FCC} \tag{B5b}$$

Insertion of Eq. (B5b) into Eq. (B3) delivers the final analytical expression for the covariance of per atom energies between the faulted and FCC distributions:

$$cov(E_a^n, E_a^{FCC}) = \langle E_a^n \, E_a^{FCC} \rangle - \langle E_a^n \rangle \langle E_a^{FCC} \rangle \tag{B6a}$$

$$\approx \frac{1}{M}\left(\sum_{X,m} \bar{C}^X \bar{E}_a^{X,n,m} \bar{E}_a^{X,FCC} - \bar{E}_a^{FCC} \sum_m \bar{E}_a^{n,m}\right) \tag{B6b}$$

This final relation appears as Eq. (10) in the main text. A numerical simulation was performed to assess the accuracy of the approximation made in the calculation of $\langle V^n/_2 \, V^{FCC}/_2 \rangle$. For this purpose, faulted topologies were computationally generated for a variety of randomly arranged solid solution single crystals with different compositions. The topologies of these crystals matched the idealized (i.e., unrelaxed) coordination environments of the faulted FCC systems considered in the GPFE landscape. The distribution of per atom energies is then calculated through an atom-by-atom examination of chemistry and pair interactions, and the application of the relevant equations of the EAM potential. The expectation values and covariance terms of Eq. (B3) are determined directly from these computational measurements. This process is similar to the initialization phase of MS calculations, where idealized crystal topologies are input into the MS software and the initial energies of atoms are computed. Numerical estimates of the covariances relevant to this study are provided in Figure B1. These numerical results are overlaid with analytical estimates from Eq. (B6b), showing excellent agreement. In addition, covariance values from MS relaxation calculations are also provided. These relaxed values also show excellent agreement with analytical estimates. The largest deviations are found in the equimolar NiCr sample, with all analytical covariance estimates showing less than a ≈ 4% difference with the relaxed data.

**FIGURES**

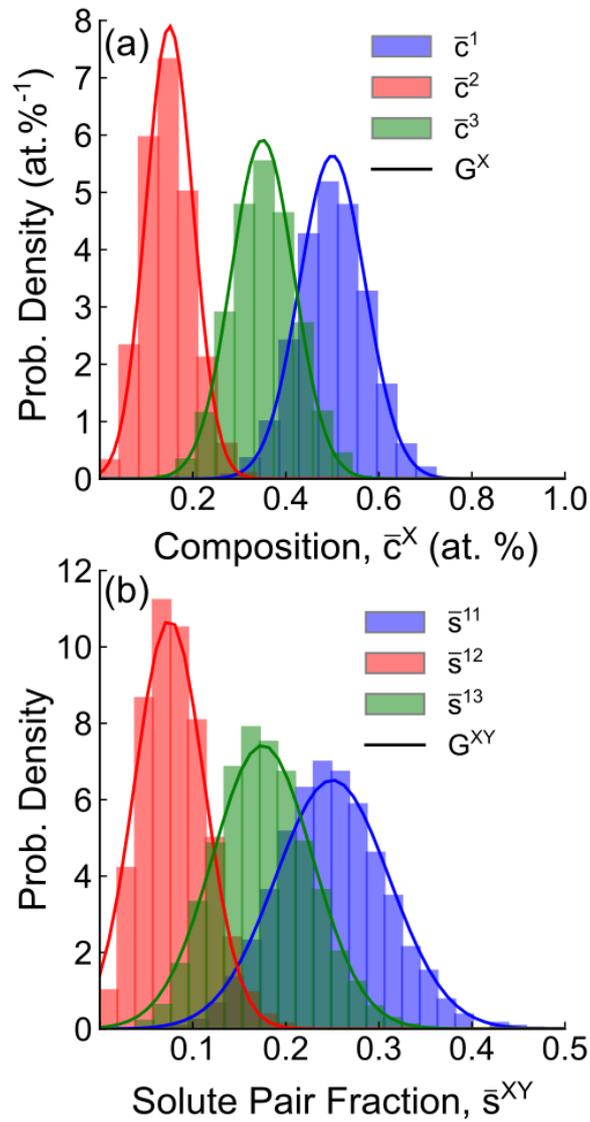

**Figure 1:** (a) A numerical simulation of sampled compositions for a randomly arranged solid solution. The sample means ($\bar{c}^X$) are plotted as a histogram and the statistical distribution ($G^X$) is plotted in solid stroke. Samples ($n_c$ = 50) were randomly collected from a ternary system of $10^6$ atoms with an overall composition of $\bar{C}$ = (0.5, 0.15, 0.35). (b) The distribution of sampled solute pair fractions ($n_s$= 50) in the same system. The sample means ($\bar{s}^{XY}$) are presented as a fraction and are plotted with the statistical distribution ($G^{XY}$) for 3 of 9 possible solute pairing, where $\bar{s}^{XY}$ and $\bar{s}^{YX}$ are separately expressed. The distributions in (a) and (b) have been sampled more than 2 x $10^4$ times with replacement.

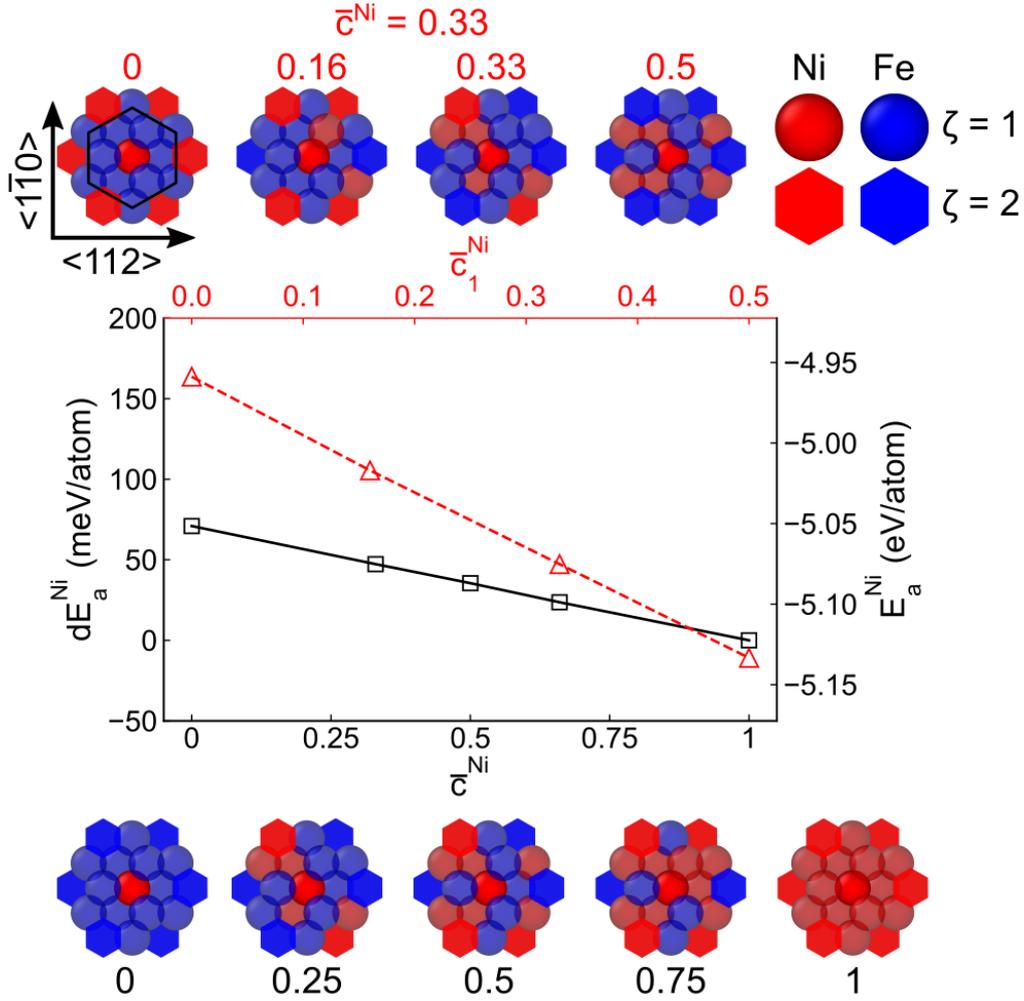

**Figure 2:** The change in per atom binding energy of a Ni atom ($dE_a^{Ni}$) in differing chemical environments within the FCC phase of a FeNi solution. For presentation simplicity, energies are calculated from interactions up to 2 nearest neighbor shells ($\zeta$) only and are referenced relative to a pure Ni sample. The right axis plots the nominal energy for the Ni atom ($E_a^{Ni}$). Here, the PE zero-point is selected at the infinite separation of atoms. The bottom axis corresponds to the sampled composition across all neighbor shells ($\bar{c}^{Ni}$). In these samples (black data points), the solute is randomly arranged in each shell but is partitioned to have an equal composition across neighbor shells (bottom schematics). The top axis plots the results for a fixed sample composition of $\bar{c}^{Ni} = 0.33$, but the composition in the first neighbor shell of the sample (i.e., $\bar{c}_1^{Ni}$) is varied. The data plotted along the top axis (red datapoints) correspond to the top schematics. The central Ni atom is not included in composition measurements. The schematics are drawn from a <111> zone axis and atoms coplanar with the central Ni atom are marked by a black hexagon for clarity. Circular and hexagonal atom markers indicate solute in the first and second neighbor shells, respectively.

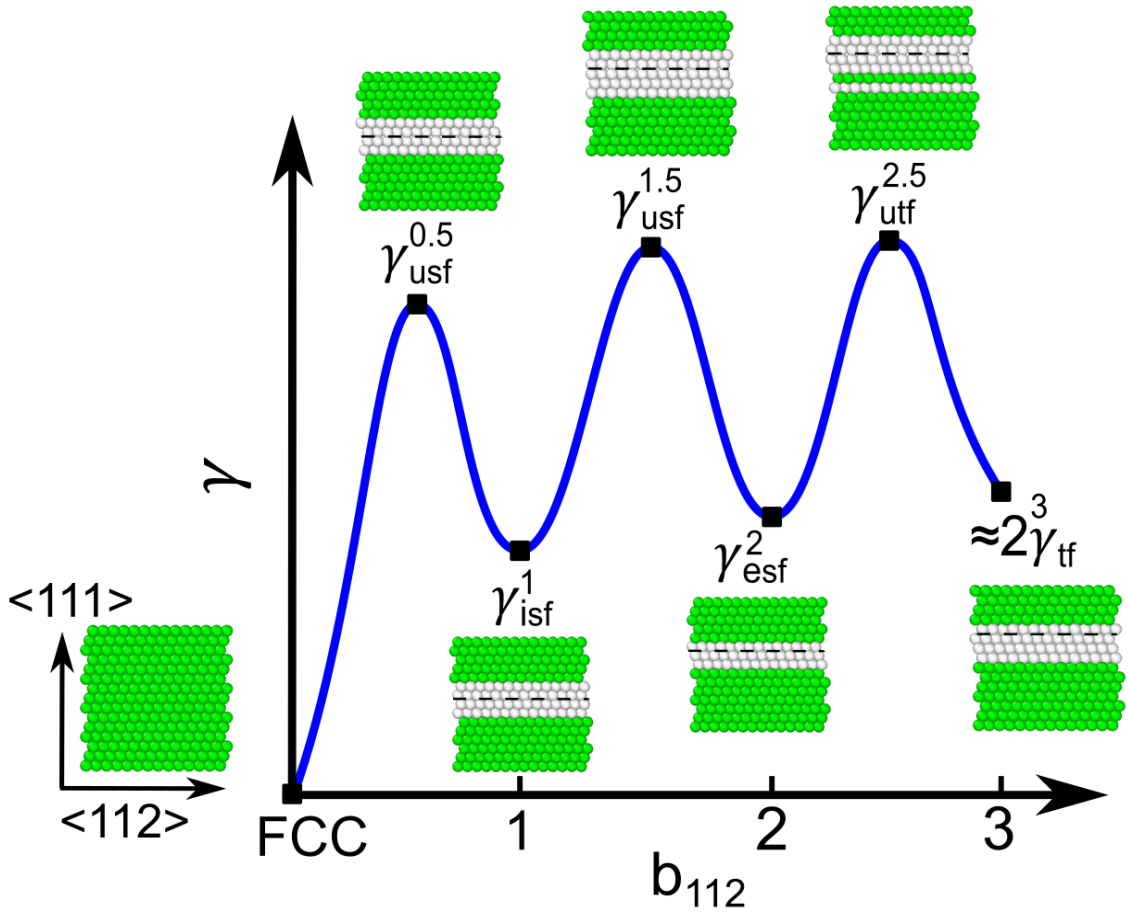

**Figure 3:** A schematic of the GPFE landscape that is formed from the interplanar excess energy of a FCC crystal. To form this curve, the crystal is sequentially sheared along <112> in consecutive {111} planes, as specified by the deformation twinning formation pathway. Shearing is measured here in increments of the Shockley partial Burger's vector ($b_{112}$). The critical energies of the GPFE landscape are indicated. $\gamma_{tf}$ is normally calculated from an isolated planar fault and is approximately equal to half the twin fault configuration shown here [47]. Snapshots of the idealized atom configurations in the vicinity of each fault are provided. Green and white atoms are classified as FCC and non-FCC coordinated, respectively. A discussion of this classification procedure is provided in the main text. Dashed lines separate the {111} crystallographic planes involved in the progressive shearing of the FCC crystal.

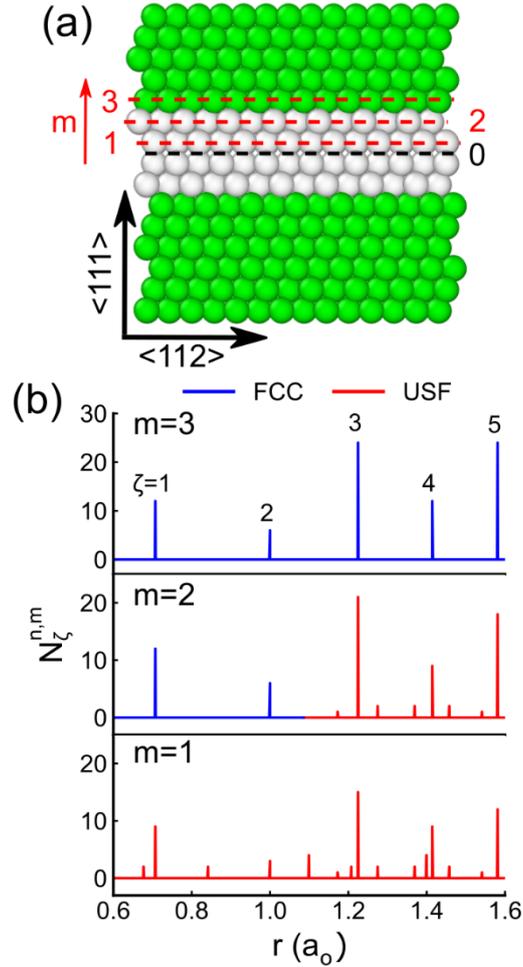

**Figure 4:** (a) The idealized configuration of atoms in the vicinity of an unstable stacking fault viewed from a <110> axis. The traces of {111} crystallographic planes (red dashed lines) are enumerated sequentially from the shearing plane ($m = 0$, black dashed line). Green and white atoms are classified as FCC and non-FCC coordinated, respectively. A discussion of this classification procedure is provided in the main text. (b) The idealized coordination relations are shown for atoms in {111} layers near the unstable stacking fault $\left(N_\zeta^{n,m} = N_\zeta^{0.5,m}\right)$. Additional coordination peaks are seen in layers adjacent to the shearing plane. The coordination relations return to the FCC sequencing at $m = 3$, as the additional coordination peaks are pushed beyond the cutoff radius. Peaks that differ in position or magnitude to the FCC coordination (blue) are colored in red. As the unstable fault is symmetric, coordination relations below the fault plane (i.e., $m = -1$, $m = -2$) are omitted. $\zeta$ is the index used to enumerate these peaks in statistical relations. An example of this process is provided for the FCC coordination ($m = 3$). Coordination relations are plotted against distance (r) as the radial position of $\zeta$ varies by defect type and {111} crystallographic plane.

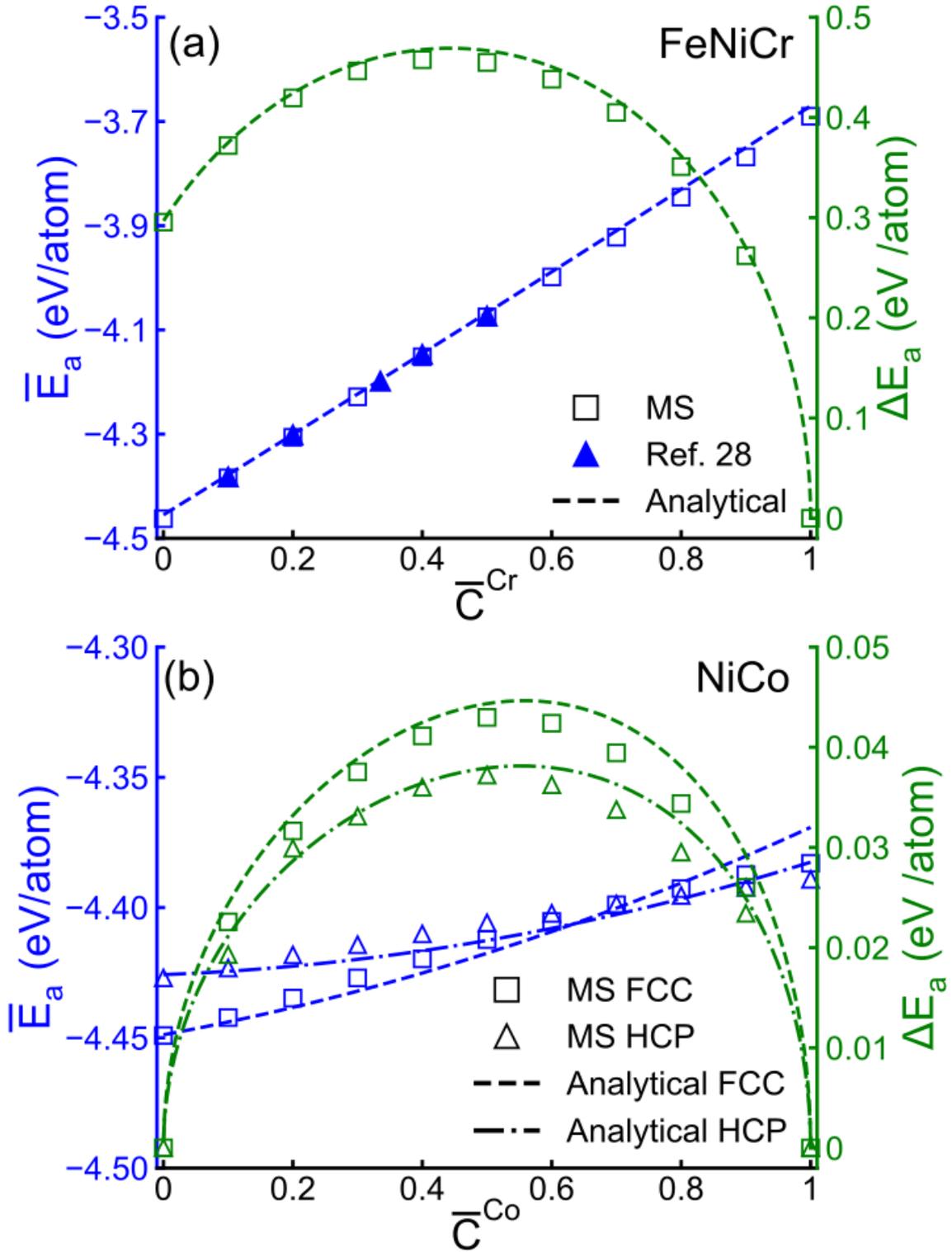

**Figure 5:** The average ($\bar{E}_a$) and standard deviations ($\Delta E_a$) of per atom binding energies in the (a) Fe$_{(1-x)/2}$Ni$_{(1-x)/2}$Cr$_x$ and (b) NiCo systems. Predictions from the statistical relations (Analytical) are overlaid with data from MS relaxation simulations, showing excellent agreement. Cohesive energy data from Varvenne et al. [28] are also provided for the FeNiCr system. The average energies and standard deviations are plotted against the primary (blue) and secondary axes (green), respectively.

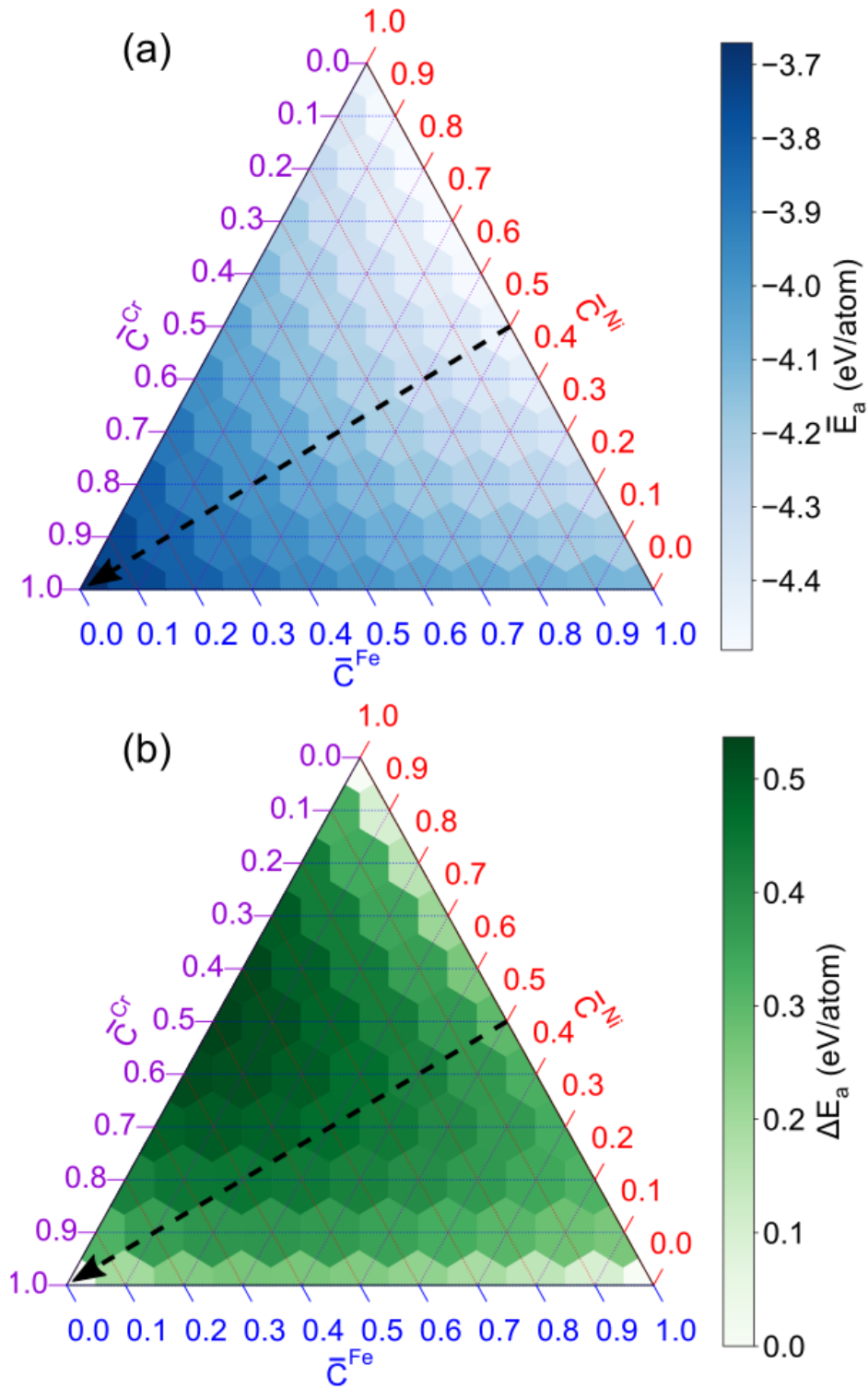

**Figure 6:** Ternary plots of the (a) average and (b) standard deviations of the per atom binding energies in the FeNiCr system. All results are calculated using the statistical relations of Section 2.1. The dashed arrow traces the data presented in Figure 5. In this system, the largest standard deviations are found along the NiCr binary terminus of this plot, and not at the equimolar composition.

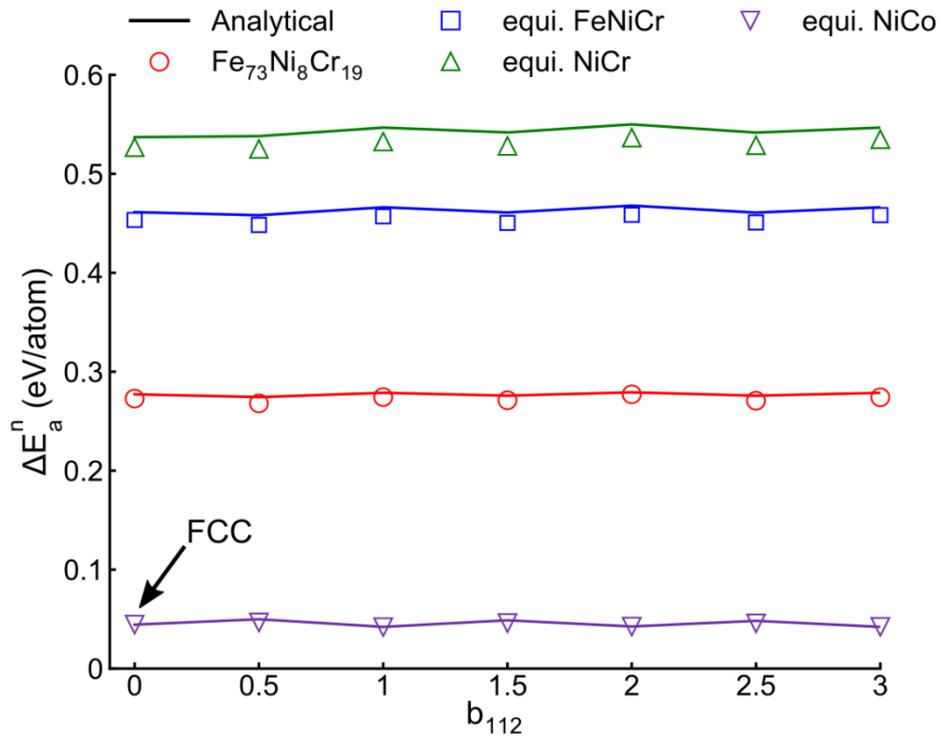

**Figure 7:** The standard deviation of per atom energies at various faulted conditions. This measurement includes contributions from {111} layers within the vicinity of planar faults. The FCC condition is shown at zero shear. MS shearing calculations are shown as markers and the predictions from statistical relations (Analytical) are overlaid in solid stroke.

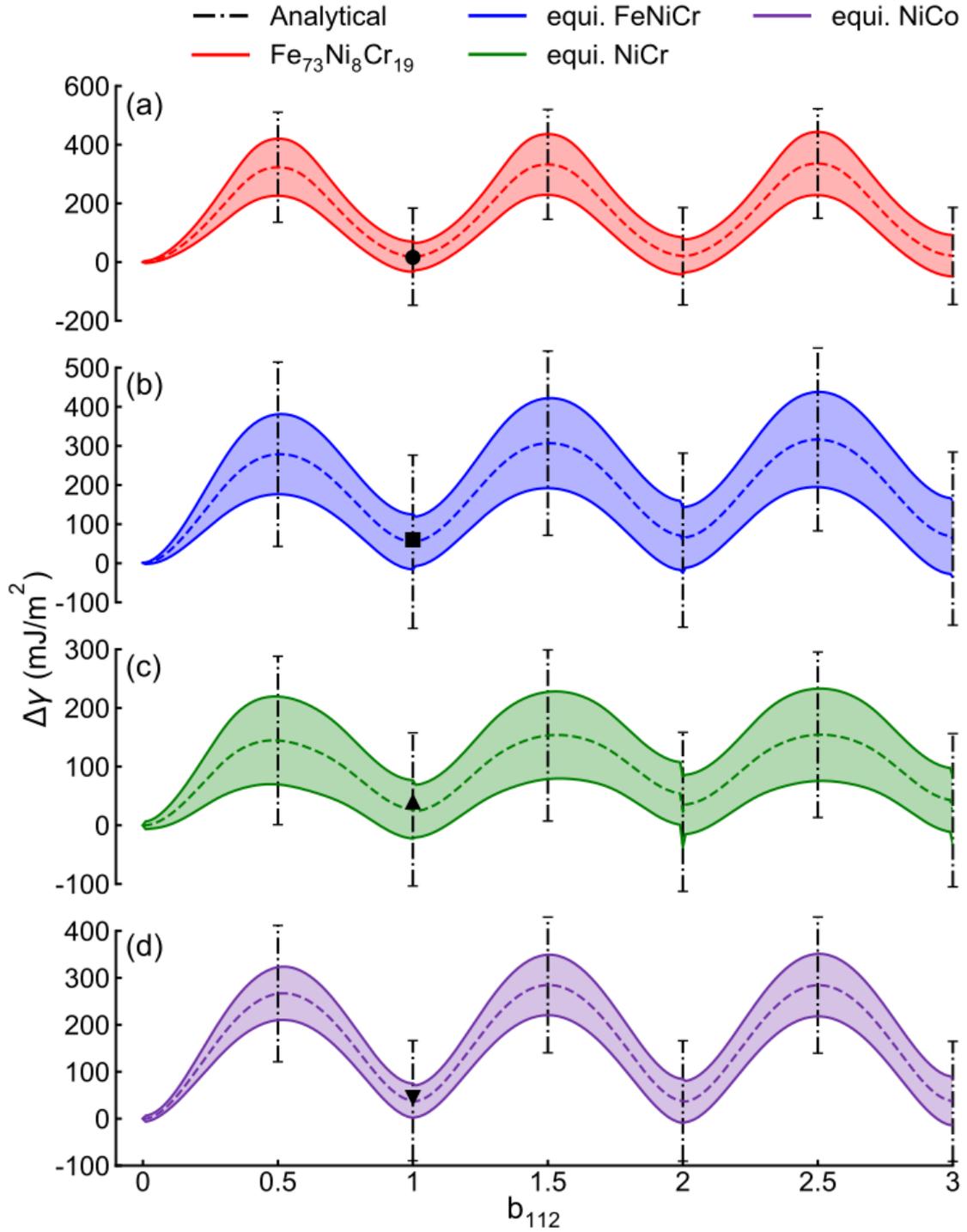

**Figure 8:** The GPFE landscapes of $Fe_{73}Ni_8Cr_{19}$ (a), and equimolar FeNiCr (b), NiCr (c) and NiCo (d). Data from MS interplanar shearing simulations is provided in the colored stroke. The average planar fault energies appear as dashed lines and a region representing ± 1 standard deviation is plotted as a filled curve bounded by solid lines. Analytical predictions for fluctuations in the critical fault energies are shown as error bars. Literature data on the average stacking fault energies of each system are provided as markers. This data is drawn from Refs. [17,28,40,67].

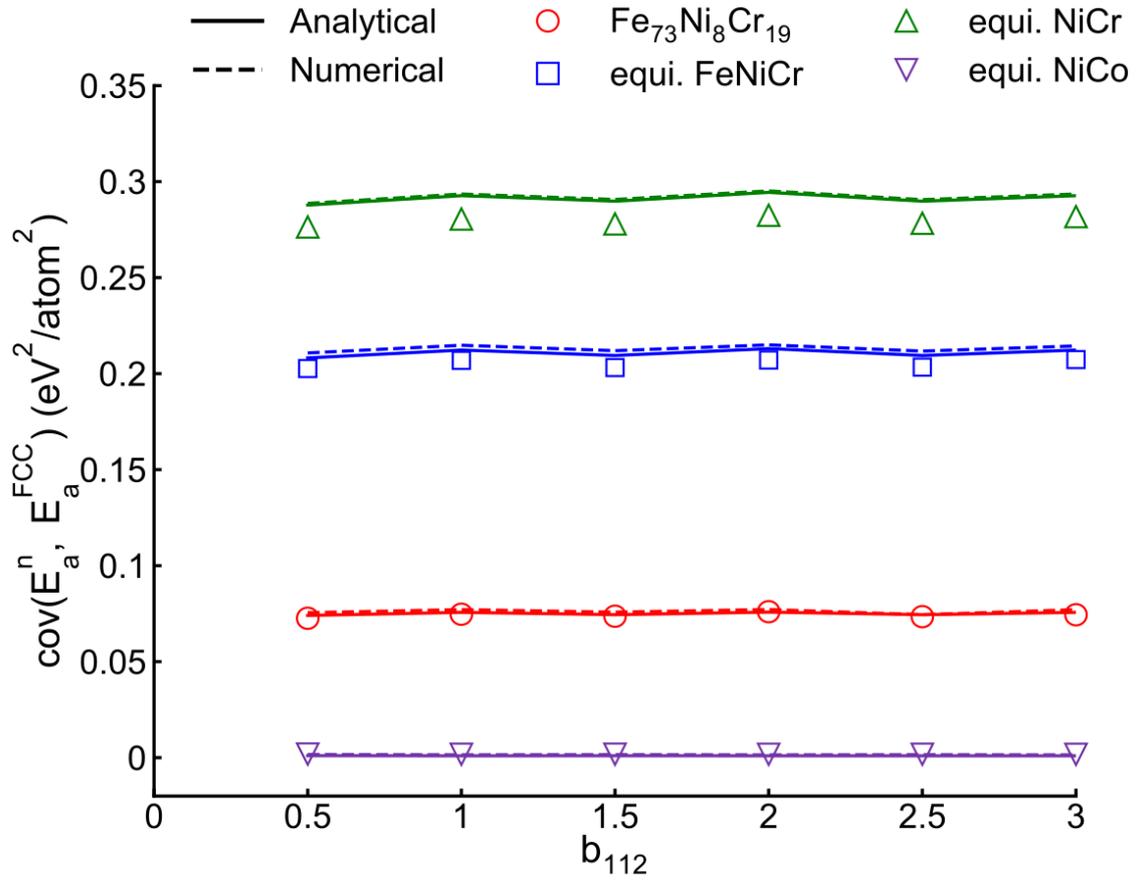

**Figure B1:** The covariance between the distributions of FCC and faulted per atom energies across different planar faults. The distribution of faulted per atom energies includes contributions from all non-FCC coordinated {111} layers within the cutoff of the planar fault. Covariance predictions from analytical (Eq. (B6b)) and numerical estimates are provided in solid and dashed stroke, respectively. Relaxed measurements from MS shearing calculations are shown as markers.